\documentclass[aps,twocolumn]{revtex4}

\usepackage{amsmath}

\begin{document}

\title{Heavy fermion behavior of itinerant frustrated systems: $\beta $-Mn,
Y(Sc)Mn$_{2}$ and LiV$_{2}$O$_{4}$}

\author{C.\ Lacroix}

\affiliation{Laboratoire Louis N\'{e}el, CNRS, BP\ 166, 38042 Grenoble Cedex 9, France}

\date{\today}

\begin{abstract}
\vspace{0.5cm}
These three metallic systems do not exhibit any magnetic ordering despite
experiments show the existence of localized moments with large
antiferromagnetic exchange: this is a consequence of the strong geometric
frustration (Y(Sc)Mn$_{2}$ and LiV$_{2}$O$_{4}$ have the Pyrochlore
structure, while \ $\beta $-Mn has a more complicated frustrated structure)
Another common feature is their very large specific heat coefficient $\gamma
=C/T$ (420 mJ mole$^{-1}$ K$^{-2}$ for LiV$_{2}$O$_{4}$)\ Several
explanations have been proposed for this ''3d heavy fermion behavior'',
including a 3d-Kondo effect. However the similarities between the three
compounds indicate that frustration plays a big role.\ We propose a new
model which takes into account the existence of two types of 3d-electrons
(localized and itinerant) and a frustrated antiferromagnetic exchange
between the localized 3d electrons.
\end{abstract}

\maketitle

\section{Introduction}

In the recent years, there has been a great interest in the study of
localized frustrated systems, but much less work whas been devoted to
itinerant frustrated systems.\ In this paper we describe three metallic
systems, Y(Sc)Mn$_2$, $\beta $-Mn and LiV$_2$O$_4$, in which the absence of
any ordered magnetic phase at low temperature is related to their frustrated
cristallographic structure.\ These three compounds exhibit many
similarities, the most spectular being their heavy Fermion behavior, wichh
is quite unusual for 3d-electron systems. Table 1 summarizes some properties
of these compounds: two of them (Y(Sc)Mn$_2$ and LiV$_2$O$_4)$, (ref 1, 2)
are similar to pyrochlore systems, while $\beta $-Mn has a more complicated
structure made of corner-sharing triangles (ref 3,4) for which we have shown
that the mean-field ground state is disordered (ref 4).\ The values of the
specific heat coefficient $\gamma =C/T$ (see 2nd line of table 1) are the
largest ones observed for 3d-metallic systems (ref 3, 5 and 6), while the
values of $\gamma $ estimated from band structure calculations is at least
one order of magnitude smaller (3th line of table 1). In all these compounds
the exchange is antiferromagnetic, but in Y(Sc)Mn$_2$, $\Theta _p$ is too
high to be measured (line 4 of table 1); another common feature is the
finite relaxation rate observed at low temperature in neutron or NMR
experiments, similar to 4f heavy fermions (ref 3, 7, 8), from which a
caracteristic spin-fluctuation temperature can be estimated (last line of
table 1). Thus, the same physical picture can be applied to these three
systems: strong frustration due to the cristallographic structure prevents
any antiferromagnetic ordered ground state and large spin fluctuations are
present up to very low temperatures; we propose here that the heavy fermion
behavior is a consequence of the peculiar properties of the ground state:
due to frustration, a large spin disorder entropy is expected ta low
temperature, giving rise to a large cobtribution to the specific heat.\ 

Several models have \-been proposed for explaining these unusual properties
of LiV$_2$O$_4$:

- the density of states has been calculated by several groups (9, 10, 11).\
In a pyrochlore lattice, flat bands can be obtained within tight binding
approximation for s-bands.\ With d-bands, some bands are almost
dispersionless, but all these band calculations, the density at the Fermi
level is at lesat 20 times too small to explain the large $\gamma $-value.\
However some interesting features can be deduced from these calculations:
the valence of V ions in LiV$_2$O$_4$ is 3,5, which indicates that half of
the V ions are V$^{4+}$ (spin 1/2) and half are V$^{3+}$ (spin 1). The
3d-electrons occupy t$_{2g}$ states: 1 electron is in A$_{1g}$ almost
localized orbital, while 0.5 electrons partially fill a twicely degenerate E$%
_g$ band.

- Fujiwara et al (ref 12) proposed that LiV$_2$O$_4$ is close to a
para-ferromagnetic instability; the large $\gamma $-value would be in this
case due to paramagnons.\ However the exchange interactions are clearly
negative in the three systems.

- The possibility of a usual Kondo effect has been discussed by Anisimov et
al (ref 13): it was proposed that there is a negative exchange between the
localized A$_{1g}$ electron and the itinerant E$_g$ electrons, but since
there is a large positive intra-atomic exchange (Hund's coupling), it is
then necessary to invoke a quite large intersite negative Kondo coupling to
overcome.

It should be mentionned that only very few papers (see e.g. 14) consider the
effect of lattice frustration in an itinerant model (Hubbard, t-J, or Kondo
lattice model).\ In this paper we study a double exchange model appropiate
for LiV$_2$O$_4$, taking explicitely into account the disorder due to
frustration.

\section{A model for LiV$_2$O$_4$}

As explained above, the band structure calculations indicate that the
3d-electrons are distributed between two types of electronic states:

- 1 electron per site is localized and will be described as a localized spin
1/2, $\overrightarrow{S_i}$.\ Nearest neighbor localized spins interact
through superexchange interaction described by the hamiltonian:

\begin{equation}
H_I=-\underset{i,j}{\sum }I_{ij}\overrightarrow{S_i}.\overrightarrow{S_j}
\;\; (I_{ij}<0)
\end{equation}

- 0.5 electrons per site are itinerant.\ They will be described in tight
binding approximation by:

\begin{equation}
H_B=-\underset{i,j,\sigma }{\sum }t_{ij}c_{i\sigma }^{+}c_{j\sigma }
\end{equation}

we neglect here the degeneracy of the conduction band and the inta-atomic
Coulomb interactions between conduction electrons. In the following we do
not take into account the real cristallographic structure and we suppose a
simple band structure with one atom per unit cell.

- Finally, we take into account the ferromagnetic intra-atomic coupling
between conduction and localized electrons due to Hund's rule:

\begin{equation}
H_J=-J_0\underset{i}{\sum }\overrightarrow{S_i}.\overrightarrow{\sigma _i}
\;\;(J_0>0)
\end{equation}

Similar models are used in the literature to describe double exchange
systems like Manganites.\ However in our case the main difference comes from
the frustrated superexchange interaction $H_I$: if this interaction is
larger than double exchange, magnetic ordering cannot be stabilized due to
frustration.\ On the contrary, if double exhange is large, ferromagnetism
should occur.\ 

Thus, we consider in the following the case where the localized spins $%
\overrightarrow{S_i}$ are in a spin liquid state which we describe by two
conditions: (i) their is no local magnetization: $<\overrightarrow{S_i}>=0$
; (ii) short range spin-spin correlations play an important role: $<%
\overrightarrow{S_i}.\overrightarrow{S_j}>\neq 0$ for neighboring sites $i$
and $j$.

\section{Slave boson mean field approximation}

We treat both interaction terms $H_I$ and $H_J$ using a slave boson mean
field approximation similar to what was used in ref 15 for the Kondo lattice
model (which is similar except the fact that $J_0$ is negative in the Kondo
model):

- for the intersite term $H_I$, we introduce fermions creation and
annihilation operators, $d_{i\sigma }^{+}$ $d_{i\sigma }^{}$ instead of the
spin operator $\overrightarrow{S_i}$ and we introduce the mean field
parameter $\Gamma _{ij}=<d_{i\sigma }^{+}d_{j\sigma }^{}>$, as in ref 15.\
Then, restricting to neraest neighbor exchange, $H_I$ is replaced by:

\begin{equation}
\widetilde{H_I}=\underset{i,\delta ,\sigma }{\sum }I_{i,i+\delta }\Gamma
_{i,i+\delta }d_{i\sigma }^{+}d_{i+\delta \sigma }^{}-\underset{i,\delta }{%
\sum }I_{i,i+\delta }(\Gamma _{i,i+\delta })^2
\end{equation}

where $\delta $ connects neighboring sites in the lattice.\ This
approximation is equivalent to a pseudo hopping hamiltonian for the
localized d-electrons; the nereast neighbor spin-spin correlation is then
negative and given by:

\begin{equation}
<\overrightarrow{S_i}.\overrightarrow{S_j}>=-\frac 32(\Gamma _{ij})^2
\end{equation}

- For the intrasite correlation term, $H_J,$ the mean field approximation of
ref.\ 15 is not appropriate since in the present case $J_0$ is positive.\
Instead, we introduce here a pesudo-hybridization between opposite spin
directions: $u_i=<c_{i\uparrow }^{+}d_{i\downarrow }>$, which leads to the
following effective hamiltonian:

\begin{eqnarray}
\widetilde{H_J}= & - & \frac{J_0}4\underset{i}{\sum }u_i(d_{i\uparrow
}^{+}c_{i\downarrow }+d_{i\downarrow }^{+}c_{i\uparrow
})+u_i^{*}(c_{i\uparrow }^{+}d_{i\downarrow }+c_{i\downarrow
}^{+}d_{i\uparrow }) \nonumber \\
 & + & \frac{J_0}2\underset{i}{\sum }u_i^2
\end{eqnarray}

The local spin-spin correlations are then positive, as expected in the case
of positive intra-atomic exchange:

\begin{equation}
<\overrightarrow{S_i}.\overrightarrow{\sigma _i}>=u_i^2
\end{equation}

- Other parameters have to be introduced to satisfy the charge conservation
conditions: on each site the d-electron number should be equal to 1; this
can be satisfied by introducing a Lagrange multiplicator $\lambda _i$.\ As a
consequence of this charge conservation, the d-states will be pinned at the
Fermi level, and this will be the origin of the large specific heat
coefficient. Finally the Fermi level $E_{F\text{ }}$is determined by fixing
the number of conduction electrons (0.5 in LiV$_2$O$_4$).\ 

Since we have not taken into account the real cristallographic structure,
both conduction and localized electrons are described by one tight binding
band, respectively $\varepsilon _c$ and $\varepsilon _d$, which are related
by: $\varepsilon _d(k)=I\Gamma \varepsilon _c(k)-\lambda $.\ This relation
allows simple analytic calculations.

We restrict to uniform solutions: $\Gamma _{ij}=\Gamma $, $u_i=u$, $\lambda
_i=\lambda $.\ In this case, the picture is that of two hybridized bands,
one of them (the d-electron band) has a temperature width $W_d\approx
2z\Gamma I$ and it is pinned at the Fermi level through the parameter $%
\lambda $. The hybridization, $V_{cd}=J_0u/4$ is also temperature dependent
and it produces a strong renormalization of the density of states at the
Fermi level, similar to the Kondo lattice. If $V_{cd}$ is large enough, it
can open a gap in the density of states, but since the system is far from
half-filling, the Fermi level will not be in the gap, but it will be located
in the peak of density of states.

The temperature-dependence of $\Gamma $, $u$ and $\lambda $ can also be
calculated, as in ref 15: it will define two caracteristic temperatures: $u$
vanishes at $T_{HF}$ which can be considered as the caracteristic
temperature below which Heavy-fermion behavior occurs ($\approx $ 20${%
{}^{\circ }}K$ in LiV$_2$O$_4$ as indicated by experiments); $\Gamma $
vanishes at $T_{mag}$ which is found to be higher than $T_{HF}$; $T_{mag}$
is related to the paramagnetic Curie temperature $\Theta _p$ ($\approx $ 60${%
{}^{\circ }}K$ in LiV$_2$O$_4$). As usual in the mean field slave boson
methods, a phase transition occurs at these caracteristic temperatures but
these transitions are detroyed if fluctuations beyond mean-field are
considered; thus, $T_{HF}$ and $T_{mag}$ should be considered as cross-over
temperatures, rather than transition temperatures.

\section{Concluding remarks}

We have proposed a new picture of the origin of the 3d-heavy heavy fermion
behavior, based on a microscopic model suitable for LiV$_2$O$_4$.\ We have
shown that a resonance may occur near the Fermi level due to the coupling of
localized and itinerant electrons in a spin liquid background: the itinerant
spins ''follow'' the localized ones through the intra-atomic coupling $J_0$
and this results in a strongly renormalized density of states for the
quasiparticles.\ There is a main difference with the Kondo case: the origin
of the Kondo resonance is in the formation of a collective singlet state due
to the Kondo interaction between one localized spin and conduction
electrons.\ Here, the singlet state is a frustration of intersite
frustration.

We have shown two different energy scales are present in this model: $T_{HF}$
and $T_{mag}$.\ These temperatures, as well as the value of $\gamma $, can
be related to the microscopic parameters $t_{ij}$, $I_{ij}$ and $J_0$; this
will be done in the near future for LiV$_2$O$_4$; estimation of these
parameters can be obtained from the band structure calculations: the
conduction band width $W_c$ is of the order of 2eV, $J_0$ is usually taken
between 0.5 eV and 1 eV.\ Estimation of the superexchange interaction $%
I_{ij} $ is more difficult since the value of the paramagnetic Curie
temperature ($\approx $-60${{}^{\circ }}K$) reflects both antiferromagnetic
superexchange and ferromagnetic double exchange.\ A crude field calculation
gives: $\Theta _p=\frac{S(S+1)}3(zI+2(J_0)^2\chi _0)$,where $\chi _0$ is the
susceptibility of itinerant electrons. A quantitative description of LiV$_2$O%
$_4$ requires also to take into account the tight binding band structure in
the spinel lattice, which contains almost flat bands already in the absence
of any correlations.\ 

Finally, in the introduction we have pointed out the similarities of the
three compounds, Y(Sc)Mn$_2$, $\beta $-Mn and LiV$_2$O$_4$, and possible
application of this model to Y(Sc)Mn$_2$ and $\beta $-Mn requires to
identify two types of electrons: localized and itinerant.

\begin{table} \centering
\begin{tabular}{c}
\begin{tabular}{lccc}
& Y(Sc)Mn$_2$ & $\beta $-Mn & LiV$_2$O$_4$ \\ 
structure & cubic Laves Phase & cubic A13 & spinel \\ 
$\gamma $ (exp.) mJ.mole$^{-1}$ K$^{-2}$ & 180 & 70 & 420 \\ 
$\gamma $ (th.) mJ.mole$^{-1}$ K$^{-2}$ & 13 & 8 & 17 \\ 
$\Theta _p$ (K) & ? & -50 & -63 \\ 
$\Gamma $ (meV) & 8 & 20 & 0.5 \\ 
T$_{sf_{}}$(K) & 160 & 400 & 10
\end{tabular}
\end{tabular}
\begin{equation*}
\end{equation*}

\caption{The three frustrated 3d-heavy fermion systems}
\end{table}

\end{document}